\documentclass[final,5p,times,number]{elsarticle}

\usepackage{graphicx}
\usepackage{amssymb}
\usepackage{natbib}
\usepackage{longtable}

\journal{Astroparticle Physics}

\makeatletter
\newcommand{\figcaption}[1]{\def\@captype{figure}\caption{#1}}
\newcommand{\tblcaption}[1]{\def\@captype{table}\caption{#1}}
\makeatother

\begin{document}

\begin{frontmatter}

\title{Search for VHE gamma rays from SS433/W50 with the CANGAROO-II telescope}

\author[konan,cor]{Sei.~Hayashi\corref{cor}}
\ead{shayashi@hep.konan-u.ac.jp}
\author[konan]{F.~Kajino}
\author[yamanashi]{T.~Naito}
\author[kyoto]{A.~Asahara}
\author[australia]{G.V.~Bicknell}
\author[adelaide]{R.W.~Clay}
\author[yamagata]{Y.~Doi}
\author[narrabri]{P.G.~Edwards}
\author[icrr]{R.~Enomoto}
\author[yamagata]{S.~Gunji}
\author[ibaraki-h]{S.~Hara}
\author[yamanashi]{T.~Hara}
\author[tokai]{T.~Hattori}
\author[ibaraki-h]{C.~Itoh}
\author[kyoto]{S.~Kabuki}
\author[hiroshima]{H.~Katagiri} 
\author[tokai]{A.~Kawachi} 
\author[icrr]{T.~Kifune}
\author[icrr]{L.T.~Ksenofontov}
\author[kyoto]{H.~Kubo}
\author[tokai]{T.~Kurihara}
\author[icrr]{R.~Kurosaka}
\author[tokai]{J.~Kushida}
\author[nagoya]{Y.~Matsubara}
\author[tokai]{Y.~Miyashita}
\author[nao]{Y.~Mizumoto}
\author[icrr]{M.~Mori}
\author[tokai]{H.~Moro}
\author[kitasato]{H.~Muraishi}
\author[konan]{Y.~Muraki}
\author[tokai]{T.~Nakase}
\author[kyoto]{D.~Nishida}
\author[tokai]{K.~Nishijima}
\author[icrr]{M.~Ohishi}
\author[icrr]{K.~Okumura}
\author[adelaide]{J.R.~Patterson}
\author[adelaide]{R.J.~Protheroe}
\author[yamagata]{N.~Sakamoto}
\author[titech]{K.~Sakurazawa}
\author[adelaide]{D.L.~Swaby}
\author[kyoto]{T.~Tanimori}
\author[kyoto]{H.~Tanimura}
\author[adelaide]{G.~Thornton}
\author[yamagata]{F.~Tokanai}
\author[nrips]{K.~Tsuchiya}
\author[icrr]{T.~Uchida}
\author[kyoto]{S.~Watanabe}
\author[konan]{T.~Yamaoka}
\author[ibaraki-u]{S.~Yanagita}
\author[ibaraki-u]{T.~Yoshida}
\author[icrr]{T.~Yoshikoshi}

\address[konan]{Department of Physics, Konan University, Kobe, Hyogo 658-8501, Japan}
\address[yamanashi]{Faculty of Management Information, Yamanashi Gakuin University, Kofu, Yamanashi 400-8575, Japan}
\address[kyoto]{Department of Physics, Graduate School of Science, Kyoto University, Sakyo-ku, Kyoto 606-8502, Japan}
\address[australia]{Research School of Astronomy and Astrophysics, Australian National University, ACT 2611, Australia}
\address[adelaide]{School of Chemistry and Physics, University of Adelaide, SA 5005, Australia}
\address[yamagata]{Department of Physics, Yamagata University, Yamagata, Yamagata 990-8560, Japan}
\address[narrabri]{Paul Wild Observatory, CSIRO Australia Telescope National Facility, Narrabri, NSW 2390, Australia}
\address[icrr]{Institute for Cosmic Ray Research, University of Tokyo,  Kashiwa, Chiba 277-8582, Japan}
\address[ibaraki-h]{Ibaraki Prefectural University of Health Sciences, Ami, Ibaraki 300-0394, Japan}
\address[tokai]{Department of Physics, Tokai University, Hiratsuka, Kanagawa 259-1292, Japan}
\address[hiroshima]{Department of Physical Science, Hiroshima University, Higashi-Hiroshima, Hiroshima 739-8526, Japan}
\address[nagoya]{Solar-Terrestrial Environment Laboratory,  Nagoya University, Nagoya, Aichi 464-8602, Japan}
\address[nao]{National Astronomical Observatory of Japan, Mitaka, Tokyo 181-8588, Japan}
\address[kitasato]{School of Allied Health Sciences, Kitasato University, Sagamihara, Kanagawa 228-8555, Japan}
\address[titech]{Department of Physics, Tokyo Institute of Technology, Meguro, Tokyo 152-8551, Japan}
\address[nrips]{National Research Institute of Police Science, Kashiwa, Chiba 277-0882, Japan}
\address[ibaraki-u]{Faculty of Science, Ibaraki University, Mito, Ibaraki 310-8512, Japan}

\begin{abstract}
SS433, 
 located at the center of the supernova remnant W50,
 is a close proximity binary system
 consisting of a compact star and a normal star.
Jets of material are directed outwards
 from the vicinity of the compact star symmetrically
 to the east and west.
Non-thermal hard X-ray emission
 is detected from lobes lying on both sides.
Shock accelerated electrons
 are expected to generate VHE gamma rays
 through the inverse-Compton process in the lobes.
Observations of the western X-ray lobe region of SS433/W50 system
 have been performed to detect VHE gamma rays using the 10\,m CANGAROO-II telescope
 in August and September, 2001, and July and September, 2002.
The total observation times
 are $85.2$~h for ON source,
 and $80.8$~h for OFF source data.
No significant excess of VHE gamma rays has been found
 at three regions of the western X-ray lobe of SS433/W50 system.
We have derived 99\% confidence level upper limits
 to the fluxes of gamma rays
 and have set constraints on the strengths of the magnetic fields
 assuming the synchrotron/inverse-Compton model
 for the wide energy range of photon spectrum from radio to TeV.
The derived lower limits are
 $4.3$\,$\mu$G for the center of the brightest X-ray emission region
 and $6.3$\,$\mu$G for the far end from SS433 in the western X-ray lobe.
In addition,
 we suggest that
 the spot-like X-ray emission may provide a major contribution
 to the hardest X-ray spectrum in the lobe.
\end{abstract}

\begin{keyword}
gamma rays: observations \sep ISM: individual (W50) \sep jets \sep stars: individual (SS433)
\PACS 95.85.Pw \sep 98.38.-j
\end{keyword}
\end{frontmatter}

\section{Introduction}

\label{sec:Introduction}

The galactic SNR W50
 is a strong non-thermal radio source.
Radio images of W50 show
 a structure extended over
 $\sim$2\,$^\circ$ $\times$ 1\,$^\circ$ 
 with limb-brightened ``ears''
 at the eastern and western ends\ \cite{Downes1986,Downes1981,Elston1987,Geldzahler1980}.
The radio emission from W50 peaks
 at 419\,mJy at 4.75\,GHz,
 with the spectral index varying
 from $\alpha$ = $0.3$ to $1.0$
 (where $S \propto \nu^{-\alpha}$)
 over the source
 in the frequency range 0.41--4.75\,GHz\ \citep{Downes1986}.
The distance to W50
 is estimated to be 5.5\,kpc\ \cite{Hjellming1981,Vermeulen1993},
 and the age is assumed to be about $10^4$\,years.
SS433, located at
 R.A.\ (J2000) = $19^h11^m49^s$,
 Dec.\ (J2000) = $+04^{\circ}58'48''$
 is the jet source located at the center of W50
 with a V-band optical magnitude of 14.2\ \cite{Margon1979}.
SS433 is a close proximity binary system
 with an orbital period 13.1\,days,
 consisting of a compact star and a normal star.
Jets of material are directed symmetrically outwards
 from the vicinity of the compact star
 to east and west
 at a speed of about $0.26c$\ \cite{Margon1989}.
The axis of the jets is precessing
 in a cone with half-angle of $20^\circ$.
The precession period is ${\sim}163$\,days,
 and the system is oriented at an angle of ${\sim}79^\circ$
 to the line of sight\ \cite{Margon1989}.
The compact star is not yet identified whether it is a black hole or neutron star.
An evaluation
 is suggested by Hillwig et al.\ (2004)\ \cite{Hillwig2004}
 that the system consists of a low-mass black hole
 with a mass of $2.9{\pm}0.7\,{\rm M}_{\odot}$
 and a type\ A3--7\ I\ supergiant
 with a mass of $10.9{\pm}3.1\,{\rm M}_{\odot}$.

Bright diffuse X-ray lobes on the 
 eastern and western sides of SS433
 were discovered by the {\it Einstein} observatory in 1983 \cite{Watson1983},
 and were confirmed by
 {\it ROSAT}~and~{\it ASCA} measurements in 1997 \cite{Safi-Harb1997}.
These X-ray lobes
 are believed to have been formed by the jets emitted from SS433,
 as they are symmetrical about SS433 and
 lie along the same axis
 as that defined by the east-west elongation of the W50 radio shell
 and are within the precession cone of the SS433 jets.
Initial {\it ROSAT}~and~{\it ASCA} measurements
 of the spectra of these X-ray lobes
 were compatible with both
 power-law model and thermal models\ \cite{Safi-Harb1997}.
Subsequently, however, the {\it ASCA} team
 reported harder power-low spectra
 of the western X-ray lobe in 2000.
The spectra were found to become softer
 with distance from SS433
 within the range of the photon index
 from ${\Gamma}=1.38$ to $2.39$\ \cite{Namiki2000a,Namiki2000b}.
They also reported that the thermal model is acceptable
 only when an unusually low metal abundance is assumed.
This indicated that
 the X-ray emissions of the western X-ray lobe have a non-thermal origin.
These results were consistent with the picture
 that the high energy electrons are generated
 at the sides of the X-ray lobes closest to SS433,
 with the electrons losing their energy by the synchrotron emission
 as they travel further from SS433.
The {\it ASCA} results suggest that
 very high energy electrons with energies up to several hundred TeV
 are expected to be produced
 through shock acceleration
 in the X-ray lobes,
 and the VHE gamma ray emissions
 are expected to be generated
 through the inverse-Compton (IC) scattering.

In 2005,
 the HEGRA team gave the flux upper limits on the VHE gamma rays
 at a few percent of the Crab nebula flux
 for the regions
 reported by the {\it ROSAT}~and~{\it ASCA} team in 1997\ \cite{Aharonian2005}.
In an effort to detect VHE gamma rays,
 we observed the western X-ray lobe region
 with the CANGAROO-II air Cherenkov imaging telescope in 2001 and 2002,
 based on the {\it ASCA} result of 2000.
We report the results of our observations
 and discuss the possible environmental conditions
 of the western X-ray lobe region.

\section{Observations}

\label{sec:Observations}

The 10\,m CANGAROO-II telescope \cite{Mori2001}
 is located near Woomera, South Australia
 ($136^{\circ}47'$E, $31^{\circ}06'$S, $160$\,m~a.s.l.)
 and consists of 114 segmented spherical mirrors
 each of 80\,cm diameter \cite{Kawachi2001}.
An imaging camera consisting of 552 PMTs is
 placed at the focal plane
 covering a field-of-view (FOV)
 of 2$^{\circ}\!\!.$76$\times$2$^{\circ}\!\!.$76.
The CANGAROO-II telescope  
 has an angular resolution of  
 0$^{\circ}\!\!.$30
 (29\,pc at a distance of 5.5\,kpc)
 with an energy threshold of 850\,GeV
 for a Crab-like energy spectrum.

\begin{figure}[htbp]
 \begin{center}
  \includegraphics[width=0.9\hsize]{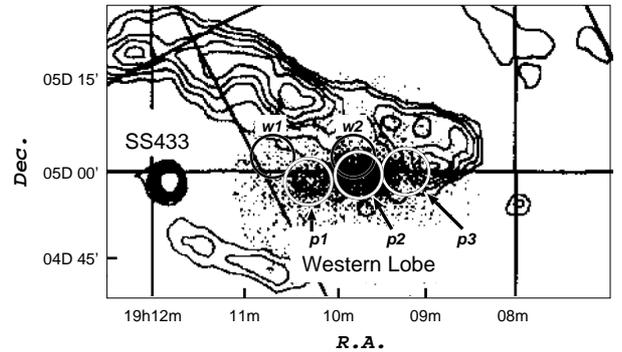}
 \end{center}
 \vspace{-0.5pc}
 \caption{%
 4.75\,GHz radio contour map of W50
 overlaid on X-ray images of the western X-ray lobe (dots)
 obtained by {\it ASCA} \cite{Namiki2000a}.
 Solid white circles show the regions
 reported by the {\it ASCA} team in 2000.
 In this paper,
 we label these regions 
 ``$p1$'', ``$p2$'' and ``$p3$''.
 Solid black circles show the regions
 reported by the {\it ROSAT}~and~{\it ASCA} team in 1997,
 which are labeled ``$w1$'' and ``$w2$''.
 }
 \label{fig:asca-map}
\end{figure}

The {\it ASCA} team reported
 the results of three regions in the western X-ray lobe in 2000,
 based on their high spatial
 and spectral resolving power over a wide energy range
 \cite{Namiki2000a,Namiki2000b}.
These regions were named
 positions 1, 2 and 3
 by the {\it ASCA} team
 in order of distance from SS433,
 centered $23'$, $31'$ and $39'$ west of SS433, respectively.
Hereafter,
 we call these regions 
 ``$p1$'', ``$p2$'' and ``$p3$'', respectively
 (Fig.~\ref{fig:asca-map}).
According to the {\it ASCA} results,
 the region ``$p1$'' shows the hardest X-ray spectrum
 of the three regions
 and has a harder X-ray spectrum than the region ``$w1$''
 which the {\it ROSAT}~and~{\it ASCA} team reported on previously.
The region ``$p2$''
 includes
 the center of the brightest X-ray region
 and ``$p3$'' is the edge of the brightest region.
The latter two regions have harder X-ray spectra than ``$w2$''.

Observations of the western X-ray lobe region were performed
 using the CANGAROO-II telescope
 in August and September 2001, and in July and September 2002.
Based on the {\it ASCA} results,
 the tracking position was set at ``$p1$'',
 since
 the region
 showed the hardest power-law spectrum with the photon index of ${\Gamma}=1.38$,
 suggesting shock acceleration could be taking place.
The data were obtained by
 ON source observations tracking the source position,
 and by OFF source observations for the background.
ON source observations were timed to
 contain the meridian passage of the target,
 as was done by Enomoto et al.\ \cite{Enomoto2002b}.
Thus, the maximum elevation angle during the observations was about 54 degrees.
OFF source observations were timed to
 have suitable offset right ascensions
 which varied day by day from $-1^h36^m48^s$ to $+4^h48^m18^s$.
The total observation times in 2001
 were 51.4~h (ON source)
 and 49.5~h (OFF source)
 and, in 2002,
 33.8~h (ON source)
 and 31.3~h (OFF source).

\section{Analysis}

\label{sec:Analysis}

We performed a preselection for the data analysis
 to obtain the data under good weather conditions.
To reduce night-sky background noise,
 we selected air shower events
 having at least 4 adjacent pixels
 with more than ${\sim}3.3$\,photoelectrons,
 which resulted in a stable shower rate.
Periods of data with a shower rate
 less than $1.5$\,Hz were not used for the present analysis
 to eliminate the effects of partial cloud
 and dew formation on the mirrors.
Moreover,
 in order to stabilize the shower rate,
 data taken at elevation angles
 less than $40^\circ$ were not used.
After these selections,
 $60.6$\% of the total observation time
 was used for the analysis (Table~\ref{tab:obstime}).
 The resulting mean elevation angle was approximately $51^\circ$.
\begin{table}[htbp]
 \begin{center}
  \caption{%
  Observation time, $t_{\rm obs}$\,hours,
  and selected time, $t_{\rm sel}$\,hours,
  in 2001 and 2002.
  }
  \label{tab:obstime}
  \begin{tabular}{crclrcl}
   \hline
   \hline
   & \multicolumn{3}{c}{ON} & \multicolumn{3}{c}{OFF} \\
   Year &
   ($t_{\rm sel}$ & / & $t_{\rm obs}$) &
   ($t_{\rm sel}$ & / & $t_{\rm obs}$) \\
   \hline
   2001 & 36.3 & / & 51.4 & 30.4 & / & 49.5 \\
   2002 & 17.2 & / & 33.8 & 16.7 & / & 31.3 \\
   \hline
   Total & 53.5 & / & 85.2 & 47.1 & / & 80.8 \\
   \hline
  \end{tabular}
 \end{center}
\end{table}

To reduce the effects of the night-sky background,
 we have used the timing information.
The pixels which were triggered
 more than $30\,nsec$ from the average trigger time of a shower
 were eliminated.
For each pixel,
 trigger counts within a $700\,{\mu}s$ period
 were recorded once a second,
 and were checked during the off-line analysis
 to exclude pixels having high trigger rates,
 which were generally caused by the passage of a bright star
 through the FOV.
Trigger counts were summed run by run
 to search for the effect of stars passing through the FOV.
After excluding pixels having more the 15 triggers per $700\,{\mu}s$ period,
 there were no apparent effects of stars
 during the whole period of observations,
 including the brightest star with the magnitude of $4.9$
 in the OFF source observations on September 6 and 7 in 2002.
Further, in each set of the data,
 we also eliminated a small number of pixels
 which showed deformed ADC spectra.
The deformed ADC spectra were determined by the following procedure.
\begin{enumerate}
 \item	The ADC spectrum of each pixel was made for both years
	using the data which satisfied the trigger condition.
 \item	The averaged ADC spectrum of 48 reference pixels
	which were located at symmetrical positions in the focal plane
	with respect to the pixel being examined
	was defined as a reference for each pixel.
 \item	${\chi}^{2}$ of the ADC spectrum
	against the corresponding reference ADC spectrum
	was obtained as ${\chi}^{2}_{\rm ADC}$ for each pixel.
	In addition,
	the number of events which satisfied the trigger condition
	was compared with averaged number for the reference pixels,
	and the ${\chi}^{2}$ of this number
	was obtained as ${\chi}^{2}_{\rm entry}$ for each pixel, too.
 \item
	Pixels which had larger values of
	${\chi}^{2}_{\rm ADC}$ and ${\chi}^{2}_{\rm entry}$
	than selected threshold values were eliminated
	in order to obtain good shower images,
	since these were possibly due to a hardware fault.
\end{enumerate}
After performing this procedure,
 the pixels having the deformed ADC spectra
 or high trigger rates
 were eliminated from the data of both years.
For each shower event,
 the lower energy events
 which have smaller size shower images
 tend to be deformed by the hardware noise.
To avoid this effect,
 we selected the events above the threshold SUMADC value
 (sum of ADC values of triggered pixels) of $2100$
 which corresponds to about $23$\,photoelectrons.

\begin{figure}[htbp]
 \begin{center}
  \begin{tabular}{cc}
   \begin{minipage}{0.45\hsize}
    \includegraphics[width=\hsize]{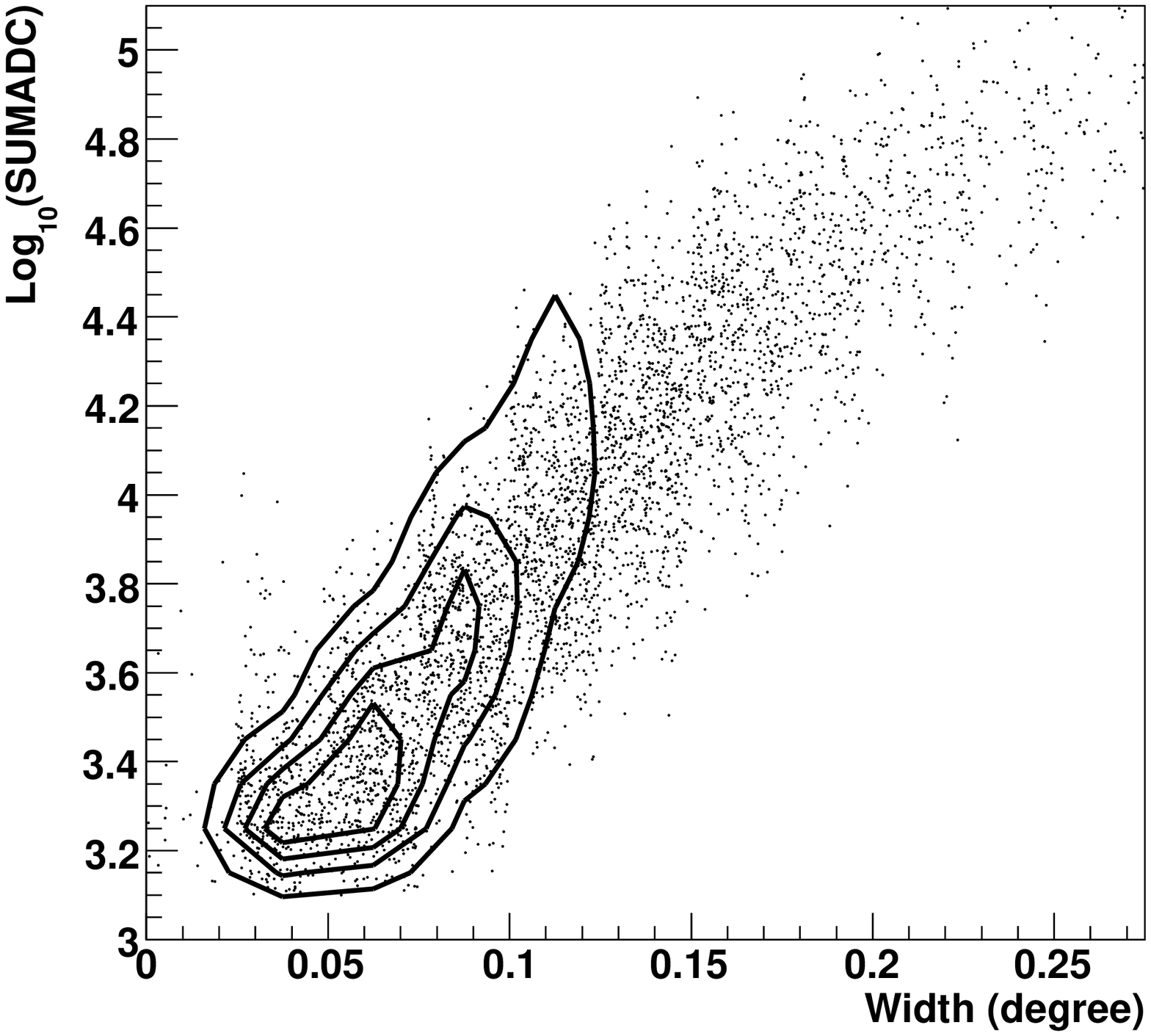}
   \end{minipage}
   \begin{minipage}{0.45\hsize}
    \includegraphics[width=\hsize]{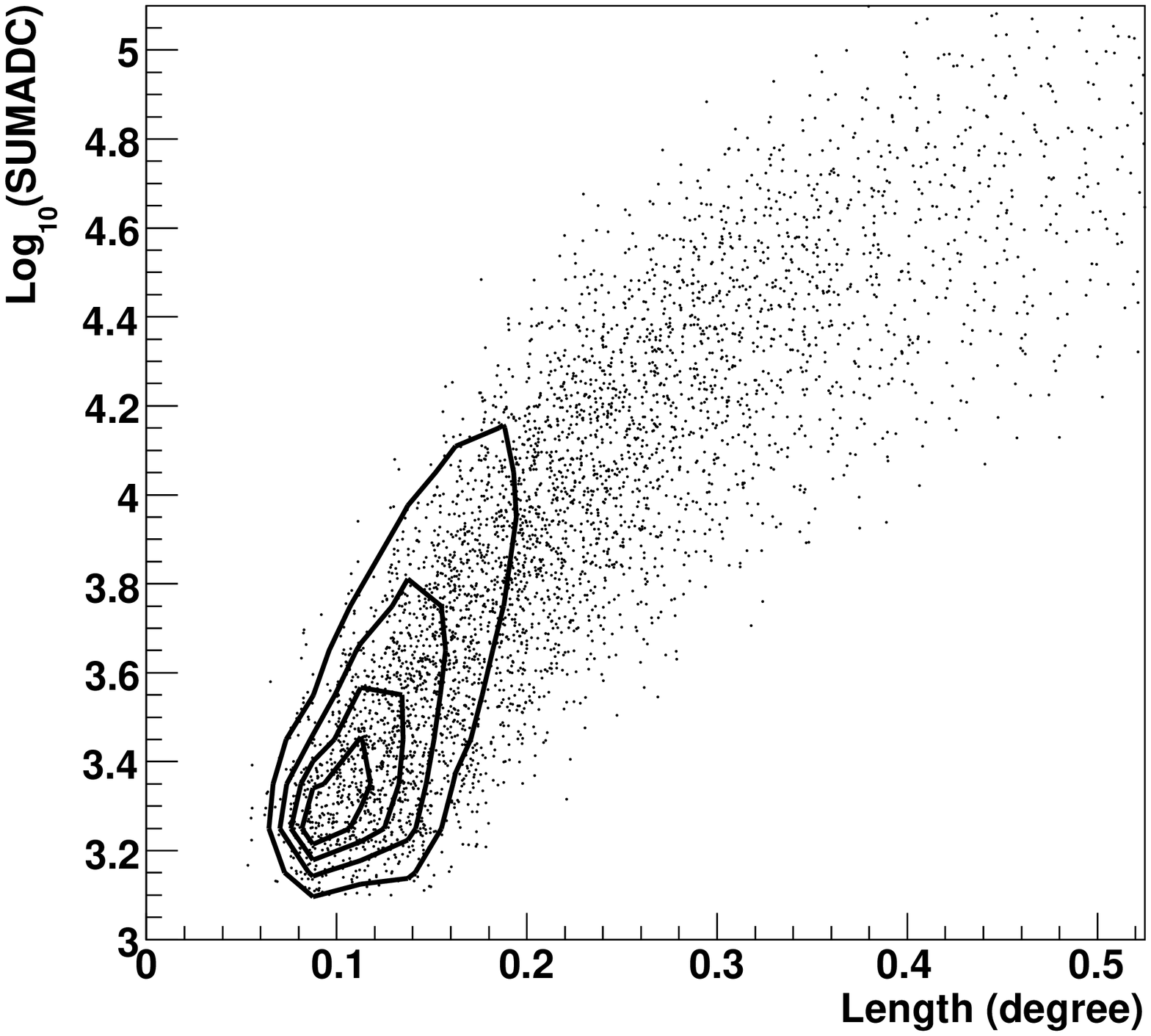}
   \end{minipage}
  \end{tabular}
 \end{center}
 \caption{%
 Distributions
 of {\it Width} (left panel)
 and {\it Length} (right panel)
 for gamma ray MC data (contours)
 and OFF-source experimental data (dots)
 in the energy range 0.2--50\,TeV.
 Each parameter depends on SUMADC
 which is the sum of ADC values event by event,
 and is approximately proportional to the energy of the incident particle.
 }
 \label{fig:width-length-dist}
\end{figure}

\begin{figure}[htbp]
 \begin{center}
  \includegraphics[width=0.6\hsize]{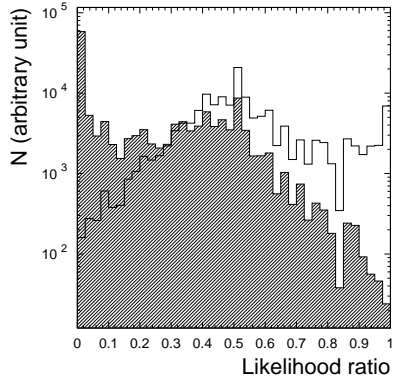}
 \end{center}
 \vspace{-0.5pc}
 \caption{%
 Distributions of Likelihood ratio ($L_{\rm ratio}$)
 for the gamma ray MC data (blank area)
 and OFF-source experimental data (hatched area).
 The number of the MC events are normalized
 to the OFF-source data.
 In this paper,
 we used the data with $L_{\rm ratio} > 0.4$.
 }
 \label{fig:likelihood-ratio}
\end{figure}

The analysis of the data was performed
 based on the imaging atmospheric Cherenkov technique\ \cite{Hillas1985,Weekes1989}.
We calculated the imaging parameters (Hillas parameters),
 {\it Distance}, {\it Length},
 {\it Width}
 using Monte Carlo (MC) simulations for gamma rays
 and OFF source data for cosmic rays.
We selected the events under the conditions:
 the distance of the centroid of the image
 from the center of the FOV was less than $1.3^\circ$
 to eliminate the edge effect of the FOV of the camera,
 and $0.2^\circ < {\it Distance} < 1.2^\circ$,
 to increase the accuracy of
 the orientation angle of the image, {\it Alpha}\ \cite{Plyasheshnikov1985,Punch1992}.
To differentiate gamma ray like events
 from cosmic ray like events,
 we adopted the Likelihood method\ \cite{Enomoto2002a}
 which has a higher selection efficiency for gamma rays
 than a conventional parameterization technique.
Figure~\ref{fig:width-length-dist}
 shows the MC results
 for the distributions of {\it Width} and {\it Length}
 of gamma rays, assuming a spectral index of $-2.5$.
The observed points for cosmic rays (background)
 are shown by the dots in the same figure.
The Likelihood method uses a single parameter:
 $L_{\rm ratio} = P_{\gamma}/\left(P_{\gamma}+P_{\rm CR}\right)$,
 where $P_{\gamma}$ and $P_{\rm CR}$ are the probability
 of the event being due to a gamma ray and a cosmic ray, respectively.
Both probabilities can be estimated from
 the products of individual probabilities
 for {\it Width} and {\it Length}
 which are derived from the probability density functions (PDFs),
 including its energy dependence.
The PDFs were obtained
 using the MC for gamma ray initiated showers
 and the OFF source data for cosmic rays.
Figure~\ref{fig:likelihood-ratio}
 shows the distributions of $L_{\rm ratio}$
 expected for gamma rays and cosmic rays.
In the region with $L_{\rm ratio} \leq 0.35$,
 cosmic rays exceed gamma rays,
 whereas in the region with
 $L_{\rm ratio} \geq 0.35$,
 gamma rays exceed cosmic rays.
In this analysis,
 we used the data with $L_{\rm ratio} > 0.4$
 to select the candidates of gamma rays.
The subsequent selection of events with $Alpha \leq 20^\circ$
 eliminates ${\sim}90$\% of the cosmic ray events
 but retains ${\sim}60$\% of the gamma ray events.

\begin{figure}[htbp]
 \begin{center}
  \begin{tabular}{c}
   \begin{minipage}{0.7\hsize}
    \includegraphics[width=\hsize]{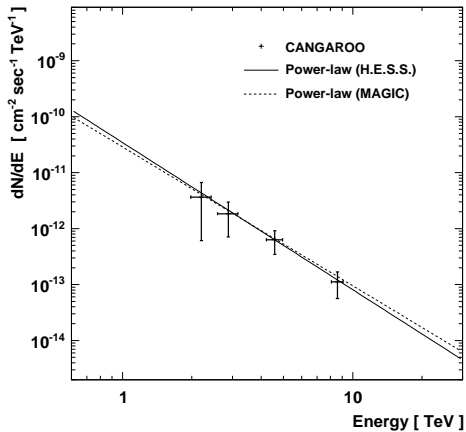}
   \end{minipage}
  \end{tabular}
 \end{center}
 \caption{Differential fluxes of Crab
 and spectra from H.E.S.S. and MAGIC.
 Only statistical errors are shown.
 }
 \label{fig:crab4}
\end{figure}

To check the feasibility of our observations and the analysis procedure,
 we analyzed Crab nebula data obtained in December 2000
 using the same analysis code.
Total analyzed times for ON source and OFF source data
 were $14.9$~h and $13.8$~h, respectively.
The maximum elevation angle was about $37$\,degrees,
 and the energy threshold was estimated to be ${\sim}2$\,TeV.
The measured gamma ray fluxes within the energy range of $2-10$\,TeV
 are shown in Table~\ref{tab:crab4} and Fig.~\ref{fig:crab4}.
Figure~\ref{fig:crab4} also shows
 power-law spectra obtained by H.E.S.S. and MAGIC\,\cite{Aharonian2006,Albert2008}.
\begin{table}[htbp]
 \caption{
 Differential fluxes of the Crab nebula.
 Only statistical errors are estimated in this analysis.
 }
 \label{tab:crab4}
  \begin{tabular}{cc}
   \hline
   \hline
   Mean energy	&	Differential flux \\
   (TeV)	&	(photons\,cm$^{-2}$\,sec$^{-1}$)	\\
   \hline
   $2.2{\pm}0.2$ & $(3.7{\pm}3.0){\times}10^{-12}$ \\
   $2.9{\pm}0.3$ & $(1.8{\pm}1.1){\times}10^{-12}$ \\
   $4.6{\pm}0.4$ & $(6.3{\pm}2.9){\times}10^{-13}$ \\
   $8.6{\pm}0.5$ & $(1.1{\pm}0.6){\times}10^{-13}$ \\
   \hline
   \hline
  \end{tabular}
\end{table}
Using the measured fluxes,
 a fit for a power-law spectrum
 gives a differential flux normalization at $1$\,TeV of
 $(2.9{\pm}2.8_{\rm stat}){\times}10^{-11}$\,cm$^{-2}$\,sec$^{-1}$\,TeV$^{-1}$
 and a power-law index of
 $2.57{\pm}0.59_{\rm stat}$.
Although the obtained spectrum
 has relatively large statistical errors,
 the differential flux of CANGAROO-II
 at $4.6$\,TeV with a power-law spectrum
 showed good agreement with
 that of H.E.S.S., within $9$\%
 and MAGIC, within $14$\%~\cite{Aharonian2006,Albert2008}.

\section{Results}

\label{sec:Results}

\begin{figure}[htbp]
 \begin{center}
  \begin{tabular}{cc}
   \begin{minipage}{0.45\hsize}
    \includegraphics[width=\hsize]{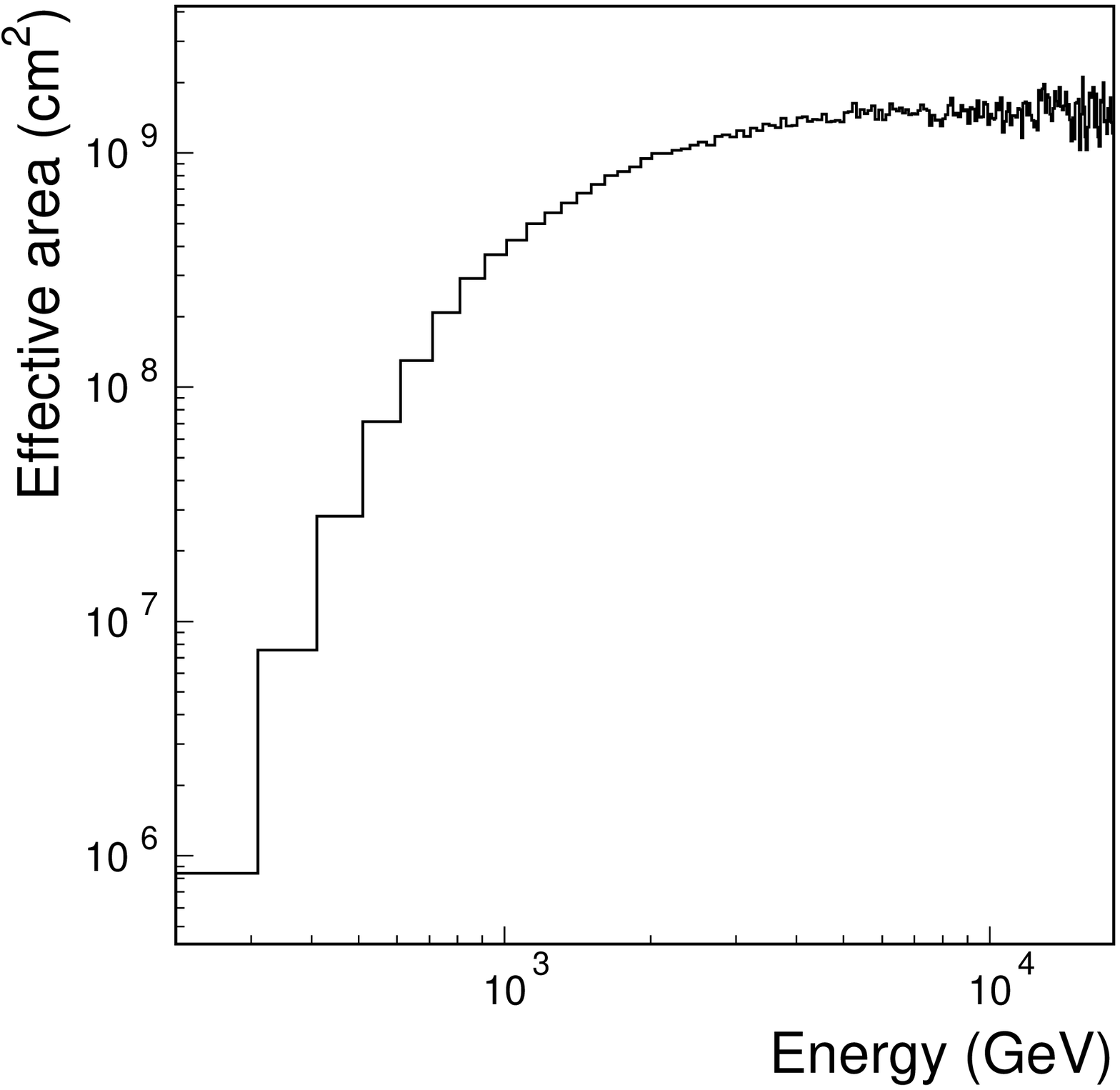}
   \end{minipage}
   \begin{minipage}{0.45\hsize}
    \includegraphics[width=\hsize]{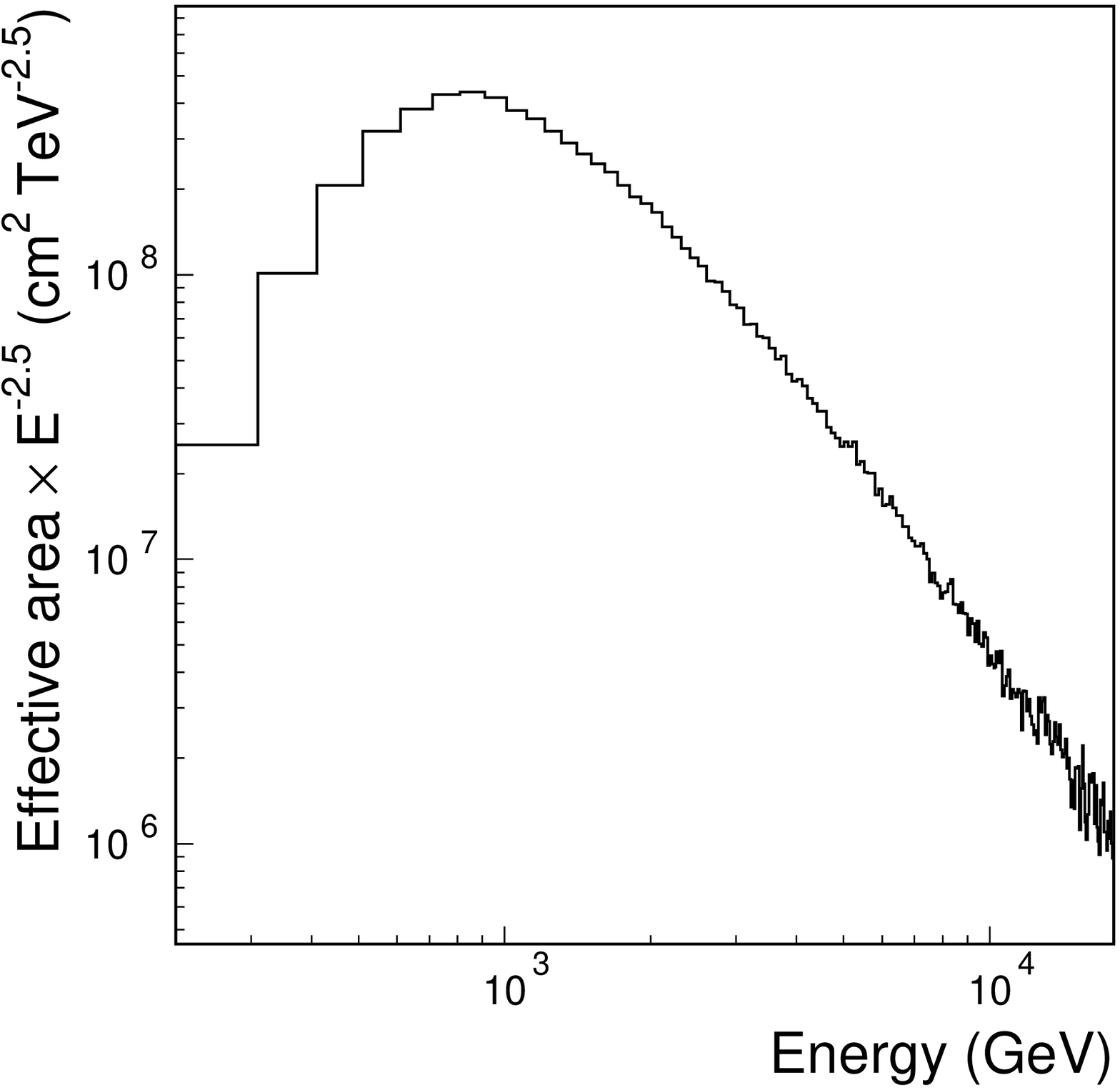}
   \end{minipage}
  \end{tabular}
 \end{center}
 \caption{%
 The effective area for gamma rays of the region ``$p1$''
 is shown in left panel.
 The right panel shows the effective area multiplied by $E^{-2.5}$.
 }
 \label{fig:pos1-seff}
\end{figure}

The left panel of Fig.~\ref{fig:pos1-seff} shows
 the energy dependence of the effective area
 for gamma rays of the region ``$p1$''
 obtained by the MC simulation.
The effective area is
 almost constant at $\sim 1.5{\times}10^{9}$\,cm$^2$ above 2\,TeV.
The right panel of Fig.~\ref{fig:pos1-seff}
 shows the energy dependence
 of the detection efficiency for gamma rays of ``$p1$''.
Thus the energy threshold for gamma rays of ``$p1$''
 is estimated to be 850\,GeV
 from this distribution.
 The effective areas and the energy thresholds for ``$p2$'' and ``$p3$''
 are estimated to be the same as for ``$p1$''.

\begin{figure}[htbp]
 \begin{center}
  \begin{tabular}{ccc}
   \begin{minipage}{0.3\hsize}
    \includegraphics[width=\hsize]{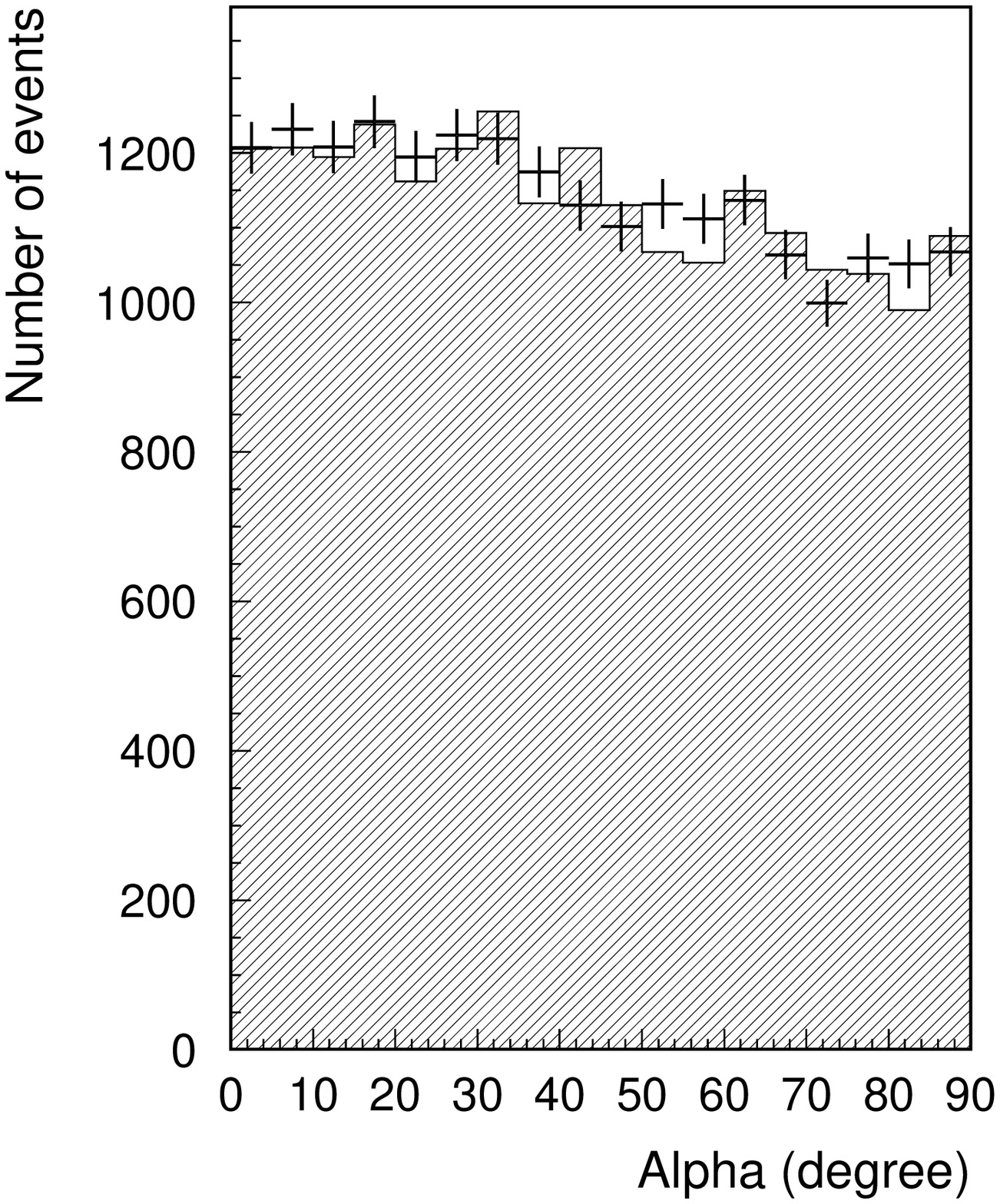}
   \end{minipage}
   \begin{minipage}{0.3\hsize}
    \includegraphics[width=\hsize]{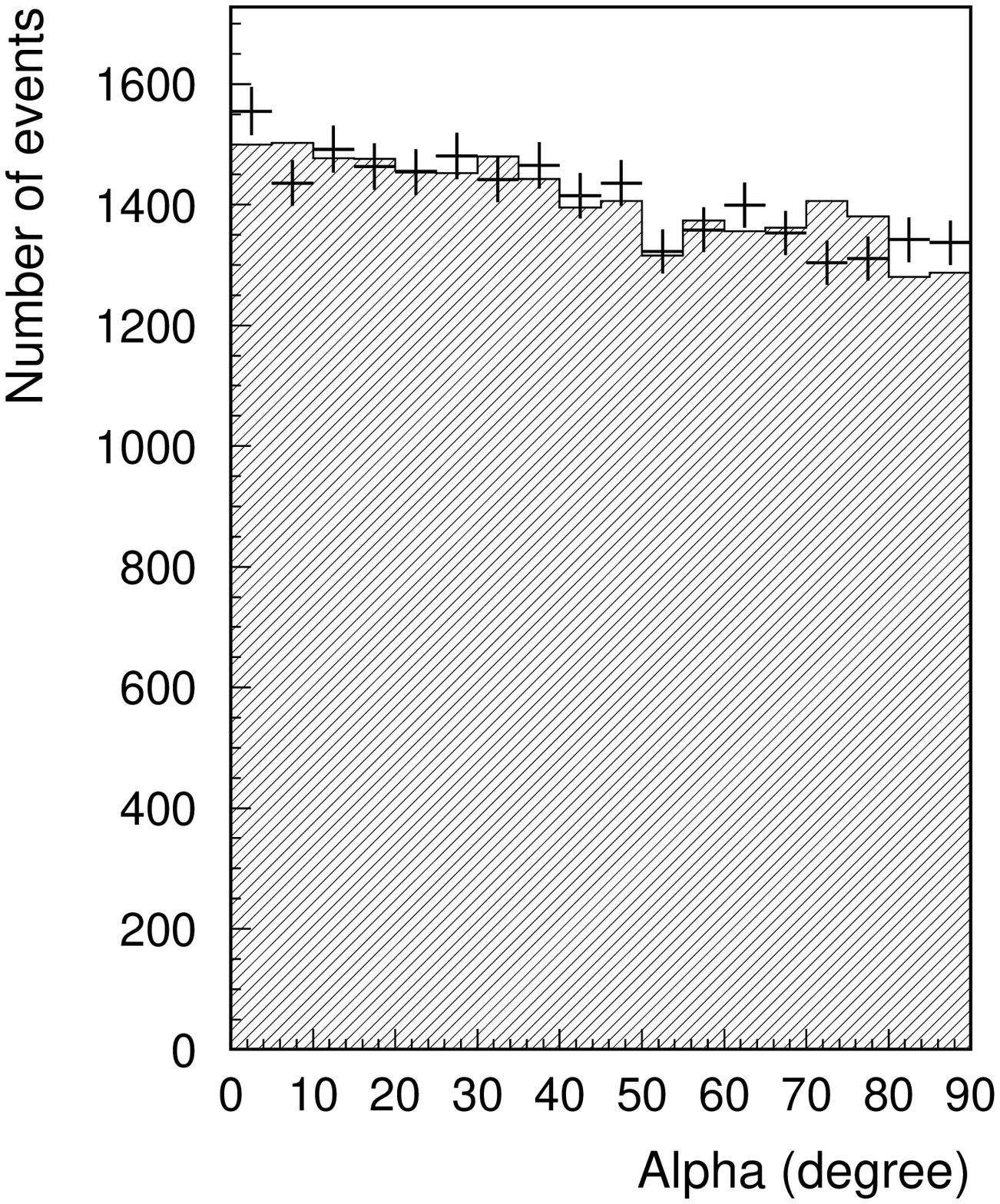}
   \end{minipage}
   \begin{minipage}{0.3\hsize}
    \includegraphics[width=\hsize]{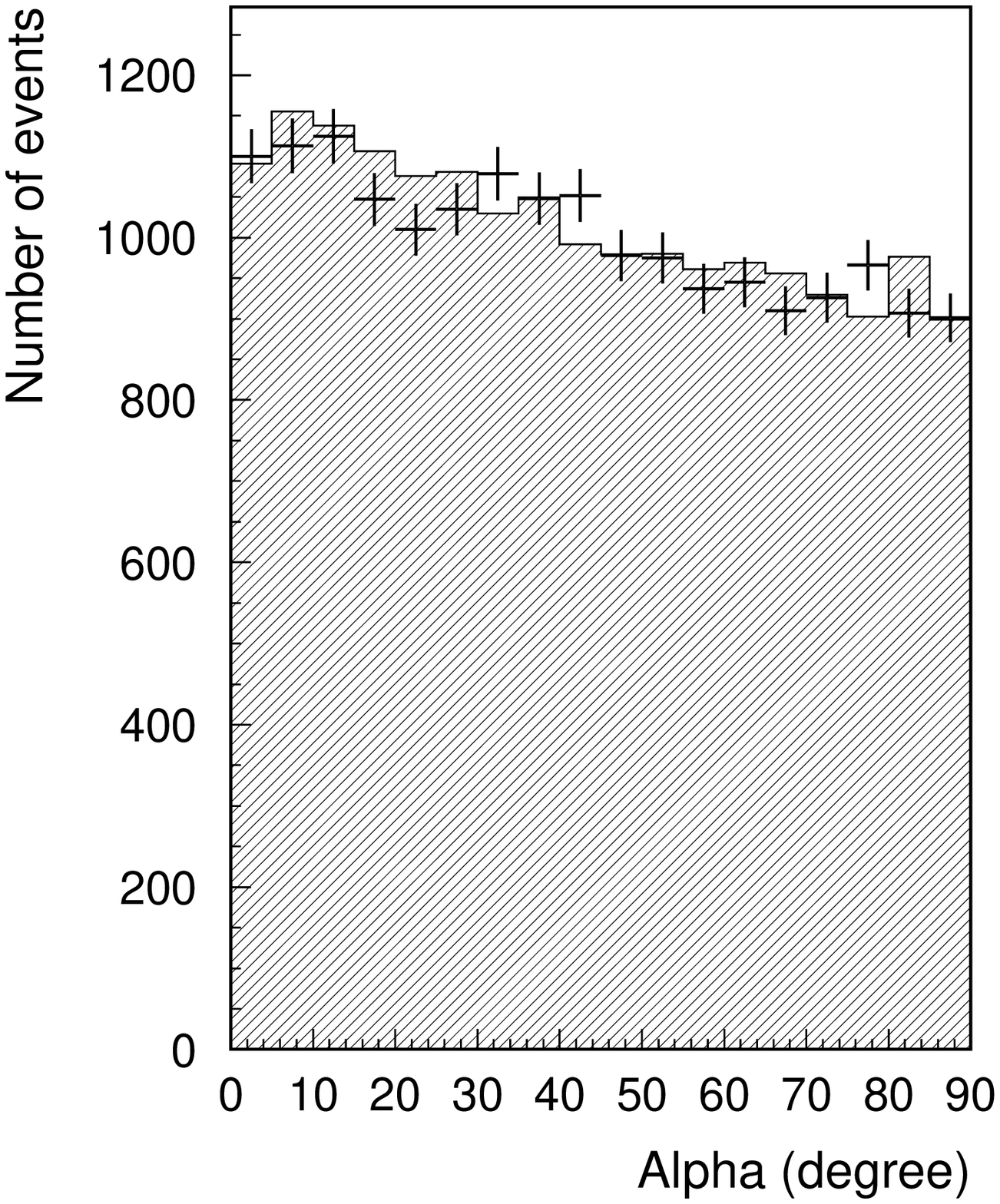}
   \end{minipage}
  \end{tabular}
 \end{center}
 \caption{%
 {\it Alpha} distributions of the combined data
 for ``$p1$'' (left), ``$p2$'' (middle) and ``$p3$'' (right).
 The points with statistical error bars show ON source data and
 hatched histograms show OFF source data.
 }
 \label{fig:pos123-alpha}
\end{figure}

Figure \ref{fig:pos123-alpha} shows
 the distributions of the image orientation angle,
 {\it Alpha}, at ``$p1$'', ``$p2$'' and ``$p3$''
 for the combined data of 2001 and 2002.
The number of OFF source events were normalized
 to the ON source data in the range of $Alpha > 30$ degrees.
The normalization factor was ${\sim}1.1$ for all regions.
The number of excess events
 was obtained
 by subtracting the number of the OFF source events
 from the ON source events
 in the range of $Alpha < 20$ degrees.

The statistical significances of the excesses
 at ``$p1$'', ``$p2$'' and ``$p3$'' were
 $0.39$, $-0.11$ and $-1.0$\,$\sigma$, respectively.
Since we found no significant excess from the data,
 we derived the 99\% confidence level upper limit fluxes,
 using Helene's method \cite{Helene1983},
 to be
 $1.5{\times}10^{-12}$,
 $1.3{\times}10^{-12}$ and
 $7.9{\times}10^{-13}$\,cm$^{-2}$\,sec$^{-1}$,
 respectively,
 for VHE gamma rays
 with $E>850$\,GeV.
These results are summarized in Table~\ref{tab:excess}.
\begin{table}[htbp]
 \caption{%
 Results of a search for VHE gamma rays
 with the CANGAROO-II telescope
 from the western X-ray lobe observed by {\it ASCA}.
 The $99\%$ C.L. upper limit fluxes are given
 above the energy threshold ($E_{\rm th} = 850$\,GeV).
 }
 \label{tab:excess}
  \begin{tabular}{cccccc}
   \hline
   \hline
   Source & R.A. & Decl. & ${^{^a}}N_s$ & ${^{^b}S}$ & $^{^c}{\phi}^{99\%}$ \\
   \hline
   $p1$
   & $19^{\rm h}10^{\rm m}17^{\rm s}$ & $+4^{\circ}57'46''$
   & 39 & 0.39 & 1.5	\\
   $p2$
   & $19^{\rm h}09^{\rm m}44^{\rm s}$ & $+4^{\circ}58'48''$
   & -12 & -0.11 & 1.3	\\
   $p3$
   & $19^{\rm h}09^{\rm m}12^{\rm s}$ & $+4^{\circ}59'13''$
   & -97 & -1.0 & 0.79	\\
   \hline
   \multicolumn{6}{l}{%
   \begin{minipage}{1.0\hsize}{\scriptsize${^a}$
    Number of Excess events included in $Alpha < 20^{\circ}$.
   }\end{minipage}}\\
   \multicolumn{6}{l}{%
   \begin{minipage}{1.0\hsize}{\scriptsize${^b}$
    Statistical significance in units of standard deviation $\sigma$\ \cite{Li-Ma1983}.
   }\end{minipage}} \\
   \multicolumn{6}{l}{%
   \begin{minipage}{1.0\hsize}{\scriptsize${^c}$
    99\% C.L. upper limit flux for $E {\geq} E_{\rm th}$
    in unit of $10^{-12}$\,ph\,cm$^{-2}$\,s$^{-1}$.
   }\end{minipage}} \\
  \end{tabular}
\end{table}

\section{Discussion}

\label{seq:Discussion}

According to the {\it ASCA} measurements,
 the X-ray lobes are considered to be generated by non-thermal emissions
 \ \cite{Namiki2000a,Namiki2000b}.
The CANGAROO-II observations were carried out
 based on this {\it ASCA} result,
 aiming to detect
 the VHE gamma rays,
 but no evidence was found for gamma ray emissions
 above $850$\,GeV.
We obtained
 upper limit fluxes which
 are useful to constrain the parameters
 of the western X-ray lobe
 based on the non-thermal model
 through a synchrotron/inverse-Compton (IC) emission process.
Therefore,
 we try to understand
 the wide energy range of photon spectrum
 from radio to TeV
 using the synchrotron/IC model.

From {\it ASCA} X-ray data \cite{Namiki2000a,Namiki2000b},
 we extracted five data points 
 in the energy region from 0.7 to 10\,keV
 by fitting to the power-law spectrum.
Similarly, we used the radio data of
 Downes et al.\ 
 at 1.7, 2.7 and 4.75\,GHz \cite{Downes1986,Downes1981}.
The radio and X-ray fluxes of ``$p1$'', ``$p2$'' and ``$p3$''
 were calculated from the intensities of respective energy ranges
 correcting the FOV to the $8'$ of {\it ASCA}.

As the seed photons for the IC process,
 we examined the possibility of IR radiation
 in addition to the cosmic microwave background (CMB).
First, we checked the IR photon field.
An upper limit to the IR flux
 is given by Band~(1987)~\cite{Band1987}
 for the optical filament region observed across ``$p2$''.
Using the upper limit flux, we have obtained
 the upper limits of respective energy densities
 as ${\sim}0.033$\,eV\,cm$^{-3}$ at $12$\,$\mu$m
 and ${\sim}0.047$\,eV\,cm$^{-3}$ at $100$\,$\mu$m.
Since we have no information on IR flux
 at ``$p1$'' and ``$p3$'',
 we assume the same upper limits for them.
The upper limit values for the interval wavelengths are interpolated
 assuming a power-law spectrum.

Next, we checked optical photon field.
Hillwig et al.\ (2004)\ \cite{Hillwig2004}
 claimed that the normal star in the SS433 system was
 a type A3-7 I supergiant
 with a mass of $10.9 \pm 3.1$\,M$_\odot$.
This implies a typical temperature and stellar radius of
 $T_{\rm s}{\sim}8000$\,$^\circ$K
 and $R_{\rm s}{\sim}40$\,$R_\odot$\ \cite{Venn1995}.
Assuming the normal star to be a blackbody radiator,
 we estimated the energy density
 to be ${\sim}0.067$\,eV\,cm$^{-3}$
 at the peak frequency of $5.0{\times}10^{14}$\,Hz.
This value of energy density negligibly contributes to the IC process
 compared to the CMB.
Further more,
 we evaluated the energy density of the optical filament,
 since ``$p2$''
 includes the optical filament in the western lobe.
From Boumis et al.\ (2007)\ \cite{Boumis2007},
 we obtained a energy density of
 ${\sim}10^{-4}$\,eV\,cm$^{-3}$ for H$\alpha$ line,
 and the same energy density levels for other emission lines.
Also from Mazeh et al.\ (1983)\ \cite{Mazeh1983},
 we obtained 
 ${\sim}10^{-3}$\,eV\,cm$^{-3}$ in N\,{\sc ii}\ $6584$\,\AA\ line.
Comparing these energy densities to
 that of CMB (${\sim} 0.26$\,eV\,cm$^{-3}$)
 coupled with the Klein--Nishina effects,
 the optical photon field provides
 almost negligible contribution to the IC process.
Therefore, for all regions
 we do not take into account
 the optical photon field
 as the seed photons of IC emission.
We also checked the possibilities that
 both the X-ray and radio photons of the western X-ray lobe
 can contribute as IC seed photons,
 and found negligibly small contributions compared to the CMB.

\begin{figure}[htbp]
 \begin{center}
  \includegraphics[width=0.9\hsize]{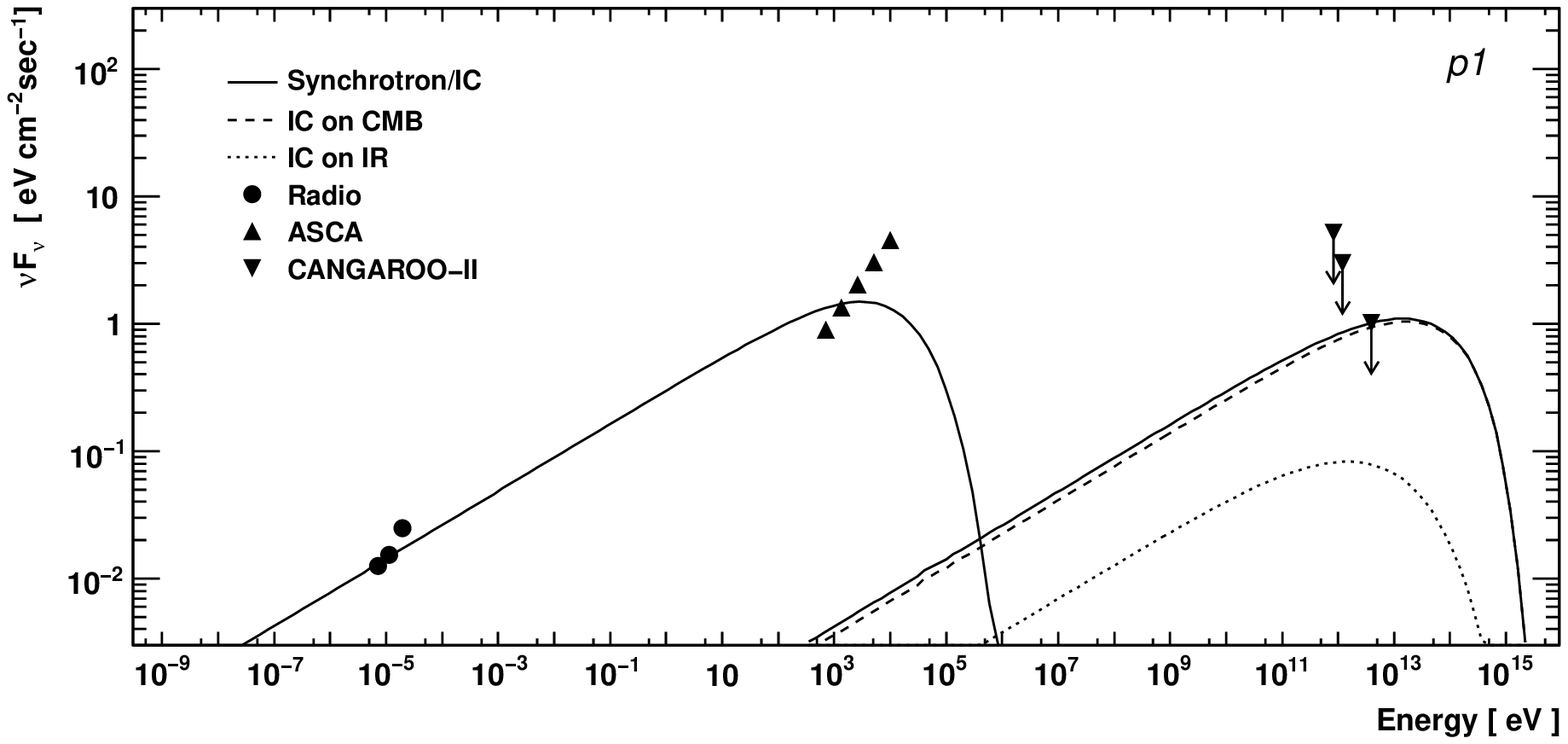}
  \includegraphics[width=0.9\hsize]{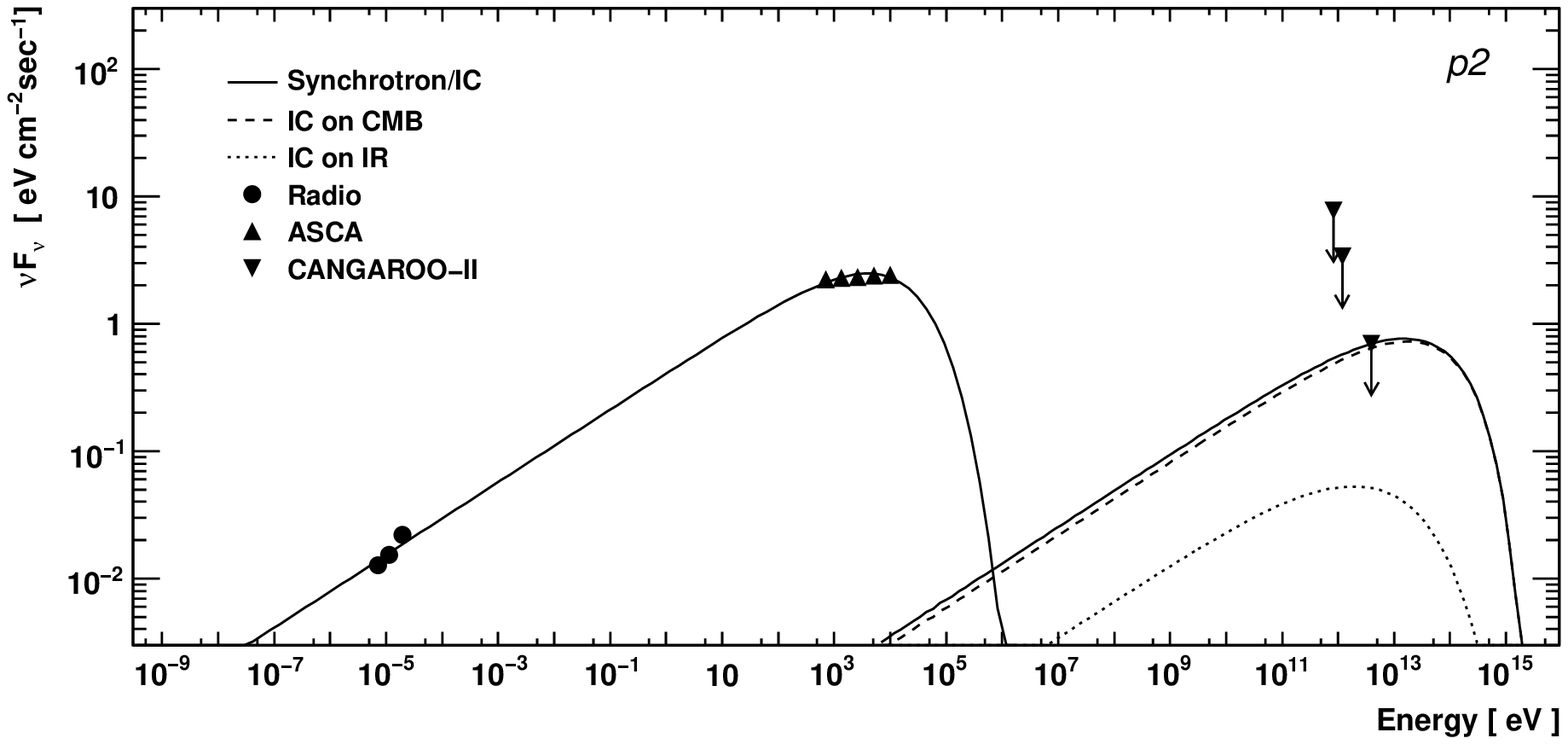}
  \includegraphics[width=0.9\hsize]{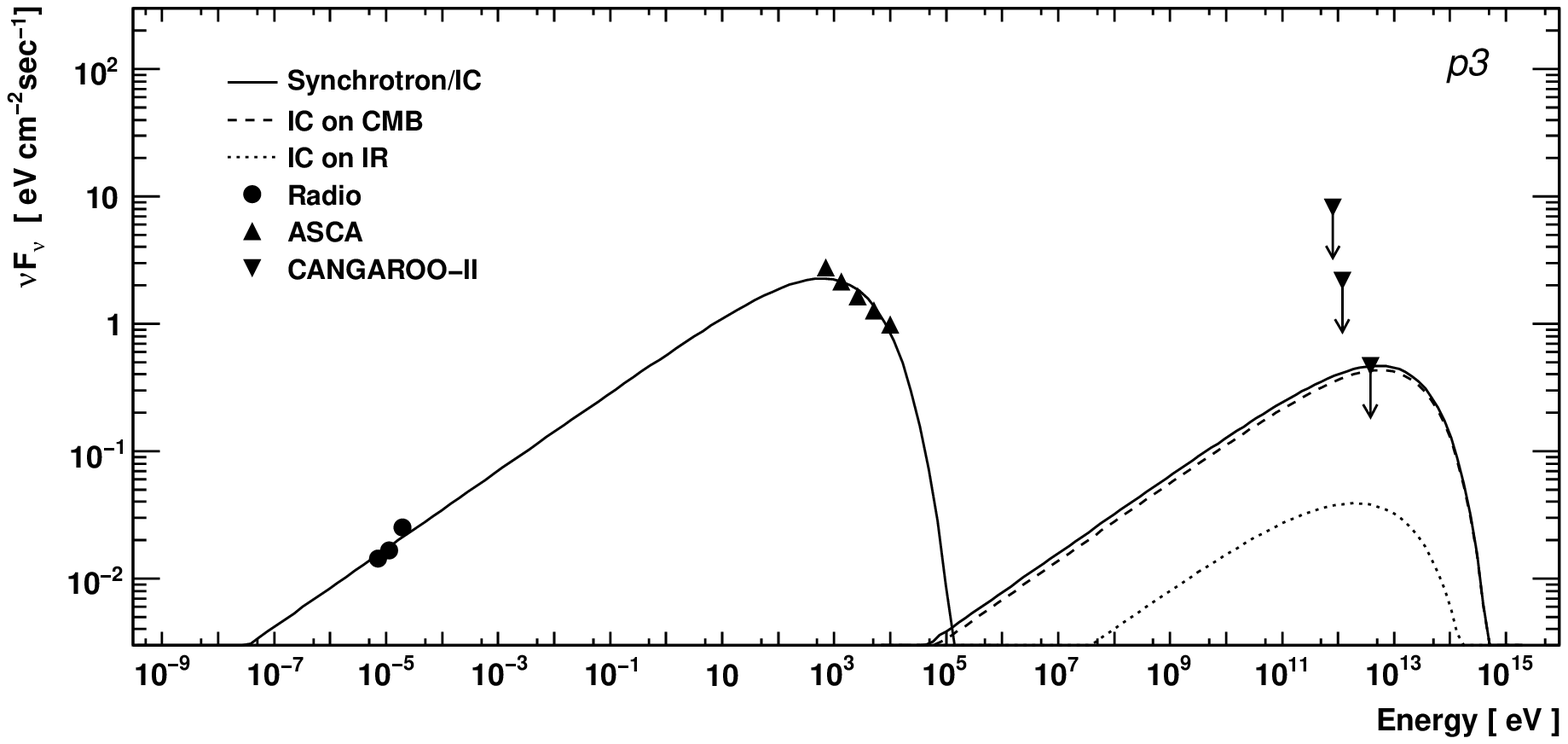}
 \end{center}
 \caption{%
 Synchrotron/IC model fitting to the spectral energy distributions at
 ``$p1$'' (top), ``$p2$'' (middle) and ``$p3$'' (bottom).
 The X-ray fluxes are from ASCA \cite{Namiki2000a,Namiki2000b} and
 the radio fluxes are calculated
 from the result of the Effelsberg 100-m telescope
 for each frequency \cite{Downes1986,Geldzahler1980}.
 The 99\% C.L. upper limit fluxes of this work are
 shown by down-arrows.
 Their mean energies are $0.9$, $1.2$ and $3.9$\,TeV, respectively.
 The thick solid lines represent the synchrotron/IC spectra.
 Two components (CMB and IR) of IC
 are shown by the dashed lines and the dotted lines,
 respectively.
 }
 \label{fig:sed-1zone}
\end{figure}

For the spectrum of electrons which drive
 the synchrotron/IC process,
 we assumed
 $E^{-\gamma}{\exp\left(-E/E_{\rm max}\right)}$
 where $\gamma$ is the power-law index,
 and $E_{\rm max}$ is
 the exponential cutoff energy of electrons.
Recent measurements of H\,{\sc i} absorption/emission spectra
 and $^{12}$CO spectrum toward SS433
 support $5.5{\pm}0.2$\,kpc
 for the distance towards the SS433/W50 system,
 and constrain the age of W50
 to be younger than $10^5$\,yr \cite{Lockman2007}.
We obtained the expected fluxes for the synchrotron/IC model
 adopting a distance of $D = 5.5$\,kpc.
The ratio between the size of synchrotron emission region
 and that of IC was assumed to be unity.
We obtained moderately good fits
 for ``$p2$'' and ``$p3$''
 on the interpretation for the wide energy range of the photon spectrum
 by freeing all four parameters.
The fit for ``$p2$'' gives
 the lower limit of the strength on the magnetic field, $B_{\rm min}$,
 to be $4.3$\,$\mu$G with ${\chi}^{2}=7.0/5$,
 and the fit for ``$p3$'' gives
 $B_{\rm min}=6.3$\,$\mu$G with ${\chi}^{2}=15/5$.
The photon indices of both regions were estimated to be ${\sim}1.7$.
For ``$p1$'', on the other hand,
 it is very difficult to understand
 the wide energy range of photon spectrum
 with a single synchrotron/IC emission model
 since the X-ray emission has a very hard spectrum.
Hence,
 we presented the result
 under the assumption of a cutoff energy of electrons
 at ``$p1$'' as $510$\,TeV
 given by
 ${\sim}280\,{\rm TeV}\,\left(E_{\rm X}/1\,{\rm keV}\right)^{1/2}\left(B/1\,{\mu}{\rm G}\right)^{-1/2}$
 ($E_{\rm X}$; the energy of X-rays generated by the synchrotron process),
 using the X-ray energy of $10$\,keV and the strength on the magnetic field of $3$\,$\mu$G.
The obtained spectral energy distributions
 for ``$p1$'', ``$p2$'' and ``$p3$''
 are shown in Fig.~\ref{fig:sed-1zone}.

\begin{figure}[htbp]
 \begin{center}
  \begin{tabular}{c}
   \begin{minipage}{0.9\hsize}
    \includegraphics[width=\hsize]{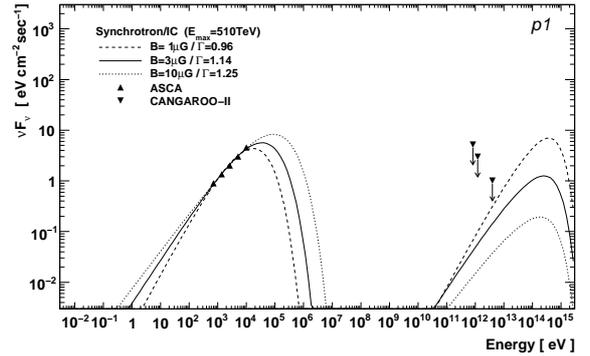}
   \end{minipage}
  \end{tabular}
 \end{center}
 \caption{%
 Alternative spectral energy distributions for ``$p1$''.
 The data shown here are the same data with Fig.~\ref{fig:sed-1zone},
 although only the X-ray data are taken into account for this fit.
 The cutoff energy of electrons is set as same value of
 $E_{\rm max}=510$\,TeV
 as Fig.~\ref{fig:sed-1zone}.
 The solid line, the dashed line and the dotted line
 represent
 the synchrotron/IC spectra
 under the various magnetic field assumptions;
 $B=1$, $3$ and $10$\,$\mu$G, respectively.
 }
 \label{fig:sed-p1}
\end{figure}

The fit for ``$p1$'' giving ${\chi}^{2}/dof=420/7$
 clearly indicates that
 a unified interpretation for the wide wavelengths from radio to TeV
 does not work well at ``$p1$''.
This result
 leads to an alternative interpretation
 that the X-ray emissions does not share
 the same radiation mechanism with the radio emissions.
According to the {\it ASCA} result,
 the radial distribution of the hard X-rays ($3.0-10.0$\,keV)
 shows a steep peak of ${\sim}3'$
 spread at near the center of ``$p1$''\ \cite{Namiki2000a}.
In contrast to such a ``hot spot'' feature,
 the radial distribution
 of the soft X-rays ($0.7-3.0$\,keV)
 does not show a clear peak
 in ``$p1$'',
 and is similar to that at ``$p2$'' and ``$p3$''
 in both energy bands.
Thus,
 the ``$p1$'' emission region is likely to be a combination
 of the spot-like hard X-ray emission at the center of ``$p1$''
 together with diffuse X-rays similar to ``$p2$'' or ``$p3$''.
A similar morphological feature has been found
 by the recent X-ray measurement with {\it XMM-Newton}
 for the eastern X-ray lobe\ \cite{Brinkmann2007}.
Based on this combined emission model for ``$p1$'',
 it can be suggested that
 the diffuse emission region shares
 the radiation mechanism from the radio to X-ray energy
 and the hard X-ray spectrum
 is mainly dominated by the spot-like emissions.
Therefore,
 we examined the synchrotron/IC emission model
 without the radio data,
 since the diffuse emissions are preferable to provide minor contribution
 for the X-ray spectrum.
The alternative spectral energy distribution for ``$p1$''
 is obtained as Fig.~\ref{fig:sed-p1},
 assuming the same cutoff energy of $E_{\rm max}=510$\,TeV
 and $B=3$\,$\mu$G
 as in Fig.~\ref{fig:sed-1zone}.
 The resulting parameters for ``$p1$'', ``$p2$'' and ``$p3$'' are 
 summarized in Table.~\ref{tab:sed-1zone}.
As shown in Fig.~\ref{fig:sed-p1},
 the expected flux of IC emissions at TeV region
 does not exceed our upper limit flux,
 hence it is found that
 the interpretation by this model
 is acceptable.
Although this model is a possible case,
 it seems to require
 the combination of multiple emission components
 to understand the wide energy range of photon spectrum
 using the synchrotron/IC model for ``$p1$''.
\begin{table}[htbp]
 \caption{%
 Fitted parameters for synchrotron/inverse-Compton model
 }
  \begin{tabular}{ccccc}
   \hline
   \hline
   &	$^{^a}{B_{\rm min}}$
   &	${E}_{\rm max}$
   &	$^{^b}{\Gamma}$
   &	${\chi}^{2}/dof$	\\
   Region
   &	(${\mu}$G)
   &	(TeV)
   &
   &			\\
   \hline
   $^{^c}$$p1$
   &	$^{^d}$3.0
   &	$^{^d}$510
   &	1.1$\pm$0.0
   &	3.5/(5-2)	\\
   $p2$
   &	4.3$\pm$0.1
   &	440$\pm$60
   &	1.7$\pm$0.0
   &	7.0/(9-4)	\\
   $p3$
   &	6.3$\pm$0.3
   &	130$\pm$10
   &	1.7$\pm$0.0
   &	15/(9-4)	\\
   \hline
   \multicolumn{5}{l}{\begin{minipage}{0.9\hsize}{\scriptsize$^a$
		       Lower limit on the strength of the magnetic field.
		      }\end{minipage}}	\\
   \multicolumn{5}{l}{\begin{minipage}{0.9\hsize}{\scriptsize$^b$
		       Photon index (=$\left(\gamma+1\right)/{2}$).
		      }\end{minipage}}	\\
   \multicolumn{5}{l}{\begin{minipage}{0.9\hsize}{\scriptsize$^c$
		       Result of a fit without the radio data.
		      }\end{minipage}}	\\
   \multicolumn{5}{l}{\begin{minipage}{0.9\hsize}{\scriptsize$^d$
		       Assumed parameters.
		      }\end{minipage}}	\\
  \end{tabular}
 \label{tab:sed-1zone}
\end{table}

\section{Conclusion}
\label{seq:Conclusion}

According to the analysis
 of the {\it ASCA} data \cite{Namiki2000a,Namiki2000b} for SS433/W50 system,
 the X-ray spectra of 3 regions in the western part of the X-ray lobe
 can be explained by non-thermal emission.
Therefore,
 shock acceleration may be present in this region
 and the emission of the VHE gamma rays may result.

Using the 10\,m CANGAROO-II telescope,
 we have searched
 for gamma rays in the VHE region
 from the western part of the X-ray lobe of SS433/W50 system.
We detected no significant excess of gamma rays
 from this region.
To check the reliability of our observations and the analysis procedure,
 we analyzed Crab nebula data using the same analysis code,
 and obtained consistent results with
 recent measurements by H.E.S.S.\ and MAGIC.
The 99\% confidence level upper limits on the fluxes of gamma rays
 for ``$p1$'', ``$p2$'' and ``$p3$'' were
 as $1.5{\times}10^{-12}$, $1.3{\times}10^{-12}$ and
 $7.9{\times}10^{-13}$\,cm$^{-2}$\,sec$^{-1}$ above $850$\,GeV, respectively.
Using these upper limit fluxes,
 we derived
 the lower limits of the magnetic field
 as to be $4.3$ and $6.3$\,$\mu$G
 for ``$p2$'' and ``$p3$'', respectively,
 under the assumption of a synchrotron/inverse-Compton model
 for the wide energy range of photon spectrum from radio to TeV.
The same interpretation for ``$p1$''
 was attempted and found to be difficult.
However,
 we suggested the alternative interpretation for ``$p1$''
 by assuming the combined X-ray emissions
 which consists of the diffused X-ray emissions
 and the spot-like hard X-ray emissions.
Since the spot-like emissions
 were supposed to
 provide major contribution
 to the hard X-ray spectrum,
 we examined the model without the radio data
 and found this interpretation was acceptable.

The authors would like to thank Dr.\ N.\ Kawai and Dr.\ M.\ Namiki
 for providing us the {\it ASCA} X-ray data and helpful comments.
This work was supported by
 a Grant-in-Aid for Scientific Research by the Japan Ministry of Education, Culture, Sports,
 Science and Technology (MEXT) of Japan,
 the 21st Century COE ``Center for Diversity and Universality in Physics'' from MEXT,
 the Australian Research Council,
 ARC Linkage Infrastructure Grant LE0238884,
 Discovery Project Grant DP0345983,
 JSPS Research Fellowships,
 and the Promotion and Mutual Aid Corporation for Private Schools of Japan.
We thank the Defense Support Center Woomera and BAE Systems.

\end{document}